\documentclass[final,times,twocolumn]{elsarticle}
\usepackage{graphicx,amssymb,amsmath}
\usepackage{multirow,dcolumn,bm,latexsym,soul,nicefrac}
\newcolumntype{K}[1]{>{\centering\arraybackslash}p{#1}}

\journal{Physics Letters B}
\begin{document}
\newcommand{\be}{\begin{equation}}
\newcommand{\ee}{\end{equation}}
\newcommand{\bq}{\begin{eqnarray}}
\newcommand{\eq}{\end{eqnarray}}

\begin{frontmatter}
\title{Scaling properties of cosmological axion strings}
\author[inst1,inst2]{C. J. A. P. Martins\corref{cor1}}\ead{Carlos.Martins@astro.up.pt}
\address[inst1]{Centro de Astrof\'{\i}sica, Universidade do Porto, Rua das Estrelas, 4150-762 Porto, Portugal}
\address[inst2]{Instituto de Astrof\'{\i}sica e Ci\^encias do Espa\c co, CAUP, Rua das Estrelas, 4150-762 Porto, Portugal}
\cortext[cor1]{Corresponding author}

\begin{abstract}
There has been recent interest in the evolution and cosmological consequences of global axionic string networks, and in particular in the issue of whether or not these networks reach the scale-invariant scaling solution that is known to exist for the simpler Goto-Nambu and Abelian-Higgs string networks. This is relevant for determining the amount and spectrum of axions they produce. We use the canonical velocity-dependent one-scale model for cosmic defect network evolution to study the evolution of these global networks, confirming the presence of deviations to scale-invariant evolution and in agreement with the most recent numerical simulations. We also quantify the cosmological impact of these corrections and discuss how the model can be used to extrapolate the results of numerical simulations, which have a limited dynamic range, to the full cosmological evolution of the networks, enabling robust predictions of their consequences. Our analysis suggests that around the QCD scale, when the global string network is expected to disappear and produce most of the axions, the number of global strings per Hubble patch should be around $\xi\sim4.2$, but also highlights the need for additional high-resolution numerical simulations.
\end{abstract}


\end{frontmatter}

\section{Introduction}

Topological defect networks are known to have a range of possible relevant cosmological roles \cite{Kibble}. In particular, global string networks can be a source of axions \cite{Peccei,Davis}. Axion strings disappear around the QCD scale ($t\sim10^{-6}$ s), as the small bias in their effective potential becomes relevant and they become attached to domain walls. However, the efficiency of axion production by global strings is an open problem, with an obvious impact on axion dark matter searches.

The main source of this uncertainty can be ascribed to limitations in numerical simulations of these networks. Global axionic strings have a logarithmically time dependent tension (or mass per unit length). This logarithm is the ratio of two length scales, one of which is cosmological (effectively the Hubble scale) while the other is microphysical (the string core width); thus one expects this logarithm to have a realistic value around 70 by the epoch when the axion strings disappear. (Note that one expects that most axions will be produced at this epoch.) On the other hand, field theory simulations need to simultaneously describe cosmological expansion and resolve the string cores, and thus computational resources imply that the range of these two scales can not be too large; in practice, only values of up to around 6 can be simulated. New numerical approaches are being tried, aiming to circumvent this limitation \cite{Klaer}. In any case, extrapolations of numerical simulations to the realistic cosmological scenario require some caution.

Continuing improvements in available computing resources, together with improvements in numerical algorithms, have nevertheless led to substantial progress in numerical simulations of global string networks. In particular, there are some indications that small---possibly logarithmic---deviations to the expected linear scaling solution \cite{Klaer,Villadoro,Kawasaki}, while there is some disagreement on what the scaling density of these networks should be. This linear scaling solution is well known from simulations of both Goto-Nambu and Abelian-Higgs strings, and for string networks in such a regime the network density will be a constant fraction of the cosmological background density. Note that in an analogous scaling regime for global strings the density will grow logarithmically with respect to the background density, since their tension is itself logarithmically time-dependent. Still, the question is whether some corrections to the linear scaling behaviour exist. Naturally, logarithmic-type corrections are difficult to confirm unambiguously in numerical simulations where the dynamic range is one of the limiting factors.

Here we discuss how analytic models can contribute to this discussion. The velocity-dependent one-scale (VOS) model is a convenient and reliable tool for studies of the evolution of networks or cosmic strings and other topological defects \cite{MS1,MS2}. Here we use it to study the evolution of axionic string networks. (An earlier, more qualitative analysis can be found in \cite{Fleury}.) We confirm that a linear scaling solution still exists but it does acquire logarithmic corrections which we quantify. We also use the model to discuss how the results of current or future numerical simulations of these networks can be extrapolated to the cosmologically realistic case.

\section{The velocity-dependent one-scale model}

The velocity-dependent one-scale (VOS) model is the canonical model for studies of the evolution of topological defect networks, in cosmological and other contexts. Here we provide a concise summary, focusing on the presently relevant case of strings in an expanding universe and referring the reader to \cite{Book} for further details.

The first assumption in this approach is to \textit{localise} the string so that we can treat it as a one-dimensional line-like object. This is clearly a good assumption for gauged strings; it is questionable for strings possessing long-range interactions, such as global strings. However, good agreement between the VOS model and simulations has been established in both local and global cases \cite{MS3}. The second assumption is to \textit{average} the microscopic string equations of motion to derive the key evolution equations for the average string velocity $v$ and correlation length $L$. These are respectively defined by
\begin{equation}
E=\mu a(\tau)\int\epsilon d\sigma\, , \qquad v^2=\frac{\int{\dot{\bf x}}^2\epsilon d\sigma}{\int\epsilon d\sigma}\, \label{eee}
\end{equation}
where $\mu$ is the string mass per unit length and $a$ is the scale factor. Bearing in mind that long string network is a Brownian random walk on large scales, the correlation length $L$ can also be used to replace the energy $E= \rho V$ in long strings in our averaged description, that is,
\be
\rho = \frac{\mu}{L^2}=\xi\frac{\mu}{t^2}\,;
\ee
in the last step we have also defined the dimensionless parameter $\xi$ which is the number of long strings per horizon, which is a convenient one for numerical simulations.

A phenomenological term must then be included to account for the loss of energy from long strings by the production of loops, leading to the evolution equation for the correlation length
\begin{equation}
2 \frac{dL}{dt} = 2HL (1+v^2)+cv\,, \label{evl0}
\end{equation}
where $H$ is the Hubble parameter and $c$ is the loop chopping efficiency. The first term in (\ref{evl0}) is due to the stretching of the network by the Hubble expansion which is modulated by the redshifting of the string velocity, while the second is the loop production term. One can also derive an evolution equation for the long string velocity with only a little more than Newton's second law
\begin{equation}
\frac{dv}{dt}=\left(1-{v^2}\right)\left[\frac{k(v)}{L}-2Hv\right]\, ,
\label{evv0}
\end{equation}
where $k$ is the \textit{momentum parameter}. The first term is the acceleration due to the curvature of the strings and the second is the damping from the Hubble expansion. Strictly speaking, the curvature radius $R$ should appear in the denominator of the first term, but in a one-scale model context one identifies $R=L$. Note that to a first approximation $k$ can be treated as constant, but in more detail an effective velocity dependence can also be identified; a detailed discussion can be found in \cite{MS2}.

Scale-invariant solutions of the form $L\propto t$ (or $L\propto H^{-1}$) together with $v=const.$, only exist when the scale factor is a power law
\begin{equation}
a(t)\propto t^\lambda\, , \qquad 0<\lambda=const. <1\, \,.
\label{conda0}
\end{equation}
By looking for stable fixed points in the VOS equations, one finds the scaling solution
\begin{equation}
\left(\frac{L}{t}\right)^2=\frac{1}{\xi}=\frac{k(k+c)}{4\lambda(1-\lambda)}\, ,\qquad v^2=\frac{1-\lambda}{\lambda}\frac{k}{k+c}\,.
\label{scalsol}
\end{equation}
In such a scaling regime the string network will be a fraction
\begin{equation}
\frac{\rho_{strings}}{\rho_{\rm crit}}= \frac{32\pi}{3k(k+c)}\frac{1-\lambda}{\lambda}\, G\mu
\end{equation}
of the universe's total energy density.

\section{Evolution of axionic strings}

The simplest VOS model presented above requires two extensions to make it applicable to the case of axionic strings. For the sake of clarity we will discuss them sequentially. This can also be justified a posteriori since each of these extensions will lead to corrections to the standard scaling behaviour.

\subsection{Time-dependent tension}

The most obvious distinguishing feature of axion strings is that their tension is time-dependent. For the purposes of the VOS model this can be written
\be
\mu(t)=\mu_0\ln{\left(\frac{L}{\delta}\right)}=\mu_0 N\,, \label{muax}
\ee
where $\delta$ is the string thickness (a microscopic parameter), and for convenience we have also defined the excess tension parameter $N$. Notice that $N$ is a slowly varying parameter. Cosmologically we expect $N\sim70$, but in standard field theory simulations the limited spatial resolution implies that only values of order a few can be simulated.

An extension of the VOS model for time-varying tensions can be found in \cite{Tension}. The evolution equation for the correlation length becomes
\begin{equation}
2\frac{dL}{dt}=2HL(1 + {v^2})+c v + \frac{L}{\mu}\frac{d\mu}{dt}\,, \label{vosmu}
\end{equation}
while the velocity equation is unchanged. In this case, substitution of Eq. (\ref{muax}) into Eq. (\ref{vosmu}) leads to
\begin{equation}
\left(2-\frac{1}{N}\right)\frac{dL}{dt}=2HL(1 + {v^2})+c v\,; \label{vosN}
\end{equation}
note that we must have $N>1$ (in other words, in this equation one can no longer recover the Goto-Nambu limit by setting $N=1$). Since $N$ is a slowly varying parameter, one can find the following implicit scaling solution for a generic expansion rate $\lambda$
\begin{equation}
\left(\frac{L}{t}\right)^2=\frac{k(k+c)}{4\lambda\left(1-\lambda-\frac{1}{2N}\right)}\, ,\qquad v^2=\frac{1-\lambda-\frac{1}{2N}}{\lambda}\frac{k}{k+c}\,.
\label{scalsol2}
\end{equation}
For the specific case of the radiation-dominated era, relevant for axion strings, we thus have
\begin{equation}
\left(\frac{L}{t}\right)^2=\frac{k(k+c)}{1-\frac{1}{N}}\, ,\qquad v^2=\left(1-\frac{1}{N}\right)\frac{k}{k+c}\,,
\label{scalax}
\end{equation}
together with
\begin{equation}
\xi=\frac{1-\frac{1}{N}}{k(k+c)}\,.
\label{scalaxxi}
\end{equation}
We therefore see how the logarithmically varying tension provides a small correction to the usual scaling.

It is often useful to compare scaling properties of axionic strings with those of standard strings. We emphasise that there is a difference between the ratios of the numbers of strings per horizon
\be
\frac{\xi_{ax}}{\xi_{GN}}=1-\frac{1}{N}
\ee
and the ratios of the physical densities of both
\be
\frac{\rho_{ax}}{\rho_{GN}}=N-1\,.
\ee
the former is less than unity, approaching it as $N\rightarrow\infty$, while the latter is larger than unity since the string tension of axionic strings is larger than the tension of a Goto-Nambu string with the same bare tension by precisely the factor $N$.

Last but not least, this analysis also shows that if one can calibrate the VOS model from simulations with a given value of $N$ it can then be used to predict the corresponding scaling parameters for other values of $N$. Specifically, from numerical measurements of the number of strings per horizon $\xi$ and the average string velocity $v$ one can obtain the VOS model parameters as follows
\be
k=\frac{v}{\sqrt{\xi}} \label{findk}
\ee
\be
c=\frac{1}{v\sqrt{\xi}}\left(1-v^2-\frac{1}{N}\right)\,. \label{findc}
\ee
It is interesting to notice that $k$ does not depend on $N$, while $c$ does.

\subsection{Radiation losses}

So far the evolution equation for the correlation length $L$ only accounts for energy losses due to loop production, whose importance is described by the loop chopping efficiency $c$. However, for global axionic strings one expects the effect of radiation backreaction losses to be more significant than that for local strings, and it should therefore be necessary to include it explicitly \cite{Battye1,Battye2}. The way to model this in the context of the VOS model has been discussed in \cite{MS2}. The result is a further term in the evolution equation for $L$, which in the notation of the present work has the form
\begin{equation}
\left(2-\frac{1}{N}\right)\frac{dL}{dt}=2HL(1 + {v^2})+c v + s \frac{v^6}{N} \,, \label{vosNS}
\end{equation}
with $s$ being a further free model parameter to be calibrated against simulations; note that this term is also $N$-dependent. The velocity equation remains unchanged.

The analysis of \cite{MS2} also shows that when this term is important but not dominant (which we expect to be the case in our present context) it can be treated perturbatively, leading to a further correction to the standard scale-invariant solution, as follows
\begin{equation}
\left(\frac{L}{t}\right)^2=\left(\frac{L}{t}\right)^2_0(1+\Delta)\, ,\qquad v^2=v_0^2(1-\Delta)\,,
\label{scalsols}
\end{equation}
where
\be
\Delta=\frac{sv_0^5}{N(k+c)}
\ee
and the index $0$ denotes the scaling parameters in the absence of radiative corrections---in our specific case, these are the ones in Eq. (\ref{scalax}). Note that this solution is now given in implicit form, since $\Delta$ itself depends on $c$, $k$ and $N$ as well as $s$. Now the number of strings per Hubble patch becomes
\be
\xi=\frac{1-\frac{1}{N}}{k(k+c)}\left[1-\frac{sv^5}{N(k+c)}\right]\,,
\ee
where we have neglected the (assumed second-order) difference between $v$ and $v_0$. Note that for the purpose of comparison with numerical simulations the parameter $k$ can still be obtained from Eq. (\ref{findk}). On the other hand there is now some degeneracy between the two parameters which model energy losses ($c$ and $s$), which is manifest in the fact that Eq. (\ref{findc}) now becomes
\be
c=\frac{1}{v\sqrt{\xi}}\left(1-v^2-\frac{1}{N}- \frac{sv^5}{N(k+c)}\right)\,. \label{findc2}
\ee
In any case, numerical methods can be used to obtain these parameters and therefore calibrate the VOS model.

\begin{table}
\begin{center}
\caption{Recent field theory simulation measurements of the number of strings per Hubble patch $\xi$ and their average velocity $v$ (if available). Note that error bars have been visually estimated from the plots in each of the given references. The excess tension parameter $N$ is also indicated in each case.}
\label{table1}
\begin{tabular}{| c | c | c | c |}
\hline
Reference & $N$ & $\xi$ & $v$ \\
\hline
{ } & $55$ & $4.4\pm0.4$ & $0.50\pm0.04$ \\
Klaer {\it et al.} \cite{Klaer} & $31$ & $4.0\pm0.4$ & $0.50\pm0.04$ \\
{ } & $15$ & $2.9\pm0.3$ & $0.51\pm0.04$ \\
\hline
Gorghetto {\it et al.} \cite{Villadoro} & $6-7$ & $1.0\pm0.3$ & N/A \\
\hline
Kawasaki {\it et al.} \cite{Kawasaki} & $2-4$ & $1.1\pm0.3$ & $0.52\pm0.05$ \\
\hline
\end{tabular}
\end{center}
\end{table}

\section{Calibration with current simulations}

We now use the extended VOS model to discuss the recent field theory numerical simulations of three different groups of authors \cite{Klaer,Villadoro,Kawasaki}. We note that these simulations differ in various ways. Apart from the obvious difference of exploring various ranges of the excess tension $N$ they also use different numerical algorithms (e.g., physical strings or fat string \cite{PRS}, standard field theory or effective field theory \cite{Klaer}) as well as different numerical diagnostics for identifying and counting the strings and (in some cases) measure their average velocities. Clearly, each of these may impact the final result. That being said, here we will take the available simulation results at face value, use them to calibrate the VOS model, and explore how the numbers of strings and their velocities depend on $N$.

Table \ref{table1} summarises the results of the three sets of simulations. We note that error bars are not statistical in the usual sense, but have been estimated---visually and somewhat conservatively--- from the plots provided in each of the given references. As for the values of $N$, for \cite{Klaer} (who simulate very different values) we list three separate sets of values, while for \cite{Villadoro,Kawasaki} (who simulate a narrower range of values) we list a single set of values for an average $N$.

\begin{figure*}
\begin{center}
\includegraphics[width=3.2in,keepaspectratio]{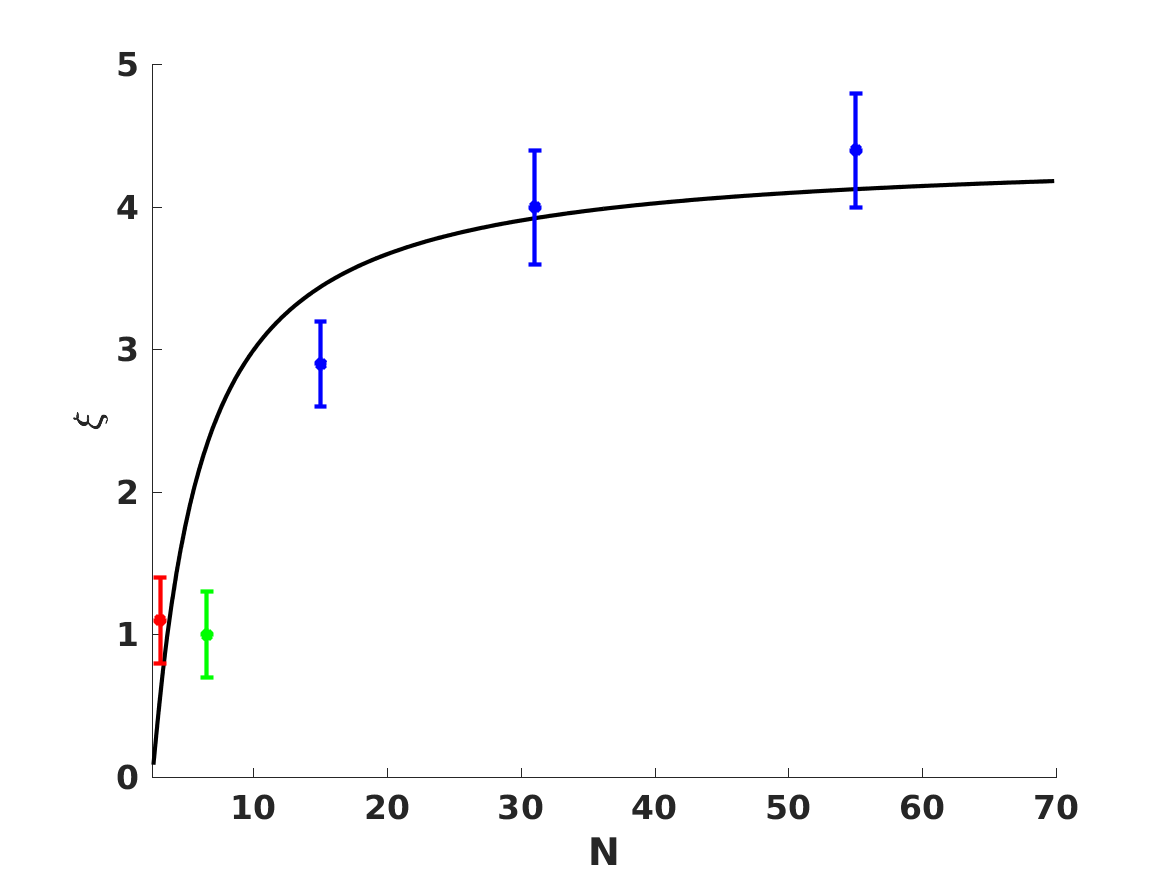}
\includegraphics[width=3.2in,keepaspectratio]{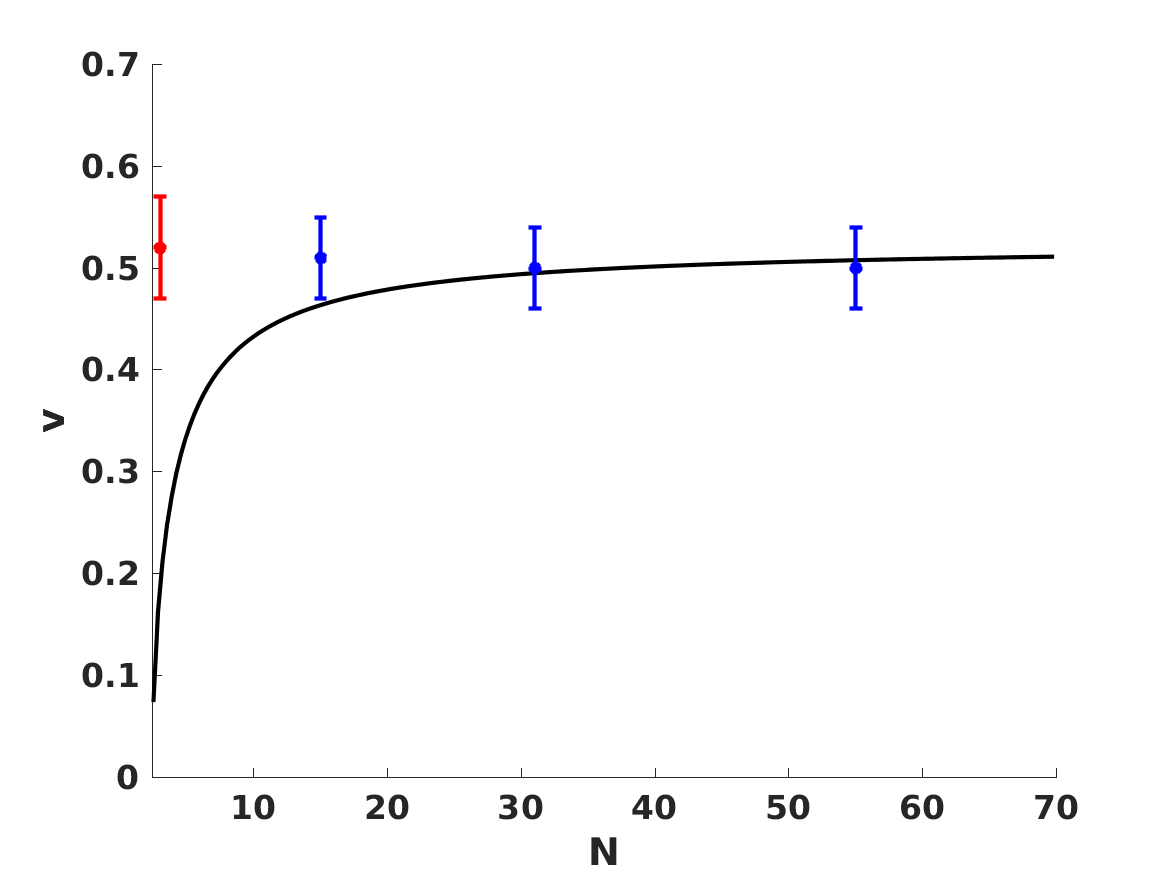}
\end{center}
\caption{\label{fig1}The dependence of the number of strings per Hubble patch (left panel) and the average string velocity (right panel) on the excess tension $N$, as predicted by the VOS model calibrated using the simulations of \cite{Klaer}. The solid line shows the model predictions, while the blue, green and red points respectively depict the results of the simulations of \cite{Klaer}, \cite{Villadoro} and \cite{Kawasaki}, as listed in Table \ref{table1}. }
\end{figure*}

We can use the three sets of simulations of \cite{Klaer} to calibrate the model. As a further simplification, noting that the string velocities do not seem to have a strong dependence on $N$ (whereas $\xi$ clearly does), we will redefine the parameter quantifying radiation losses as
\be
\sigma\equiv s v_0^5\,.
\ee
We thus estimate the following values for the VOS model parameters
\be
c\sim0.66
\ee
\be
k\sim0.25
\ee
\be
\sigma\sim2.2\,.
\ee
Since, as previously mentioned, the uncertainties listed in Table \ref{table1} are not statistical error bars, we also do not provide error bars for the model parameters. Nevertheless we have found that parameter values within about ten percent of the listed ones provide reasonable fits to the simulations of \cite{Klaer}. Note that assuming a value of $v_0\sim0.5$ this value of $\sigma$ corresponds to $s\sim70$. On the other hand, the three parameters lead to $\Delta\sim2.4/N$, implying that this calibration requires $N>2.4$; in practice this is not a problem since the cosmologically relevant values of $N$ are much larger.

Figure \ref{fig1} compares the VOS model predictions for the number of strings per Hubble patch and the average string velocity, with the aforementioned parameters (shown with the black lines), to the data in Table \ref{table1}. We see that the model indeed provides a satisfactory fit to the values of $\xi$ of both \cite{Klaer} and \cite{Kawasaki}, while that found by \cite{Villadoro} is smaller than that predicted by the best-fit model. Note that the model predicts that $\xi$ will saturate for even mild values of $N$; this is in broad agreement with the simulations of \cite{Klaer}. On the other hand \cite{Villadoro} suggests, based on simulations with $N<7$, that $\xi$ grows linearly with $N$; such a behaviour can not be reproduced in the VOS model, although it can be a reasonable approximation for small values of $N$, say $N<10$.

This saturation with $N$ also occurs for the velocities. Here we note that the VOS model would predict smaller velocities for smaller values of $N$, while the comparison of the simulations of \cite{Klaer} and \cite{Kawasaki} offers some very mild evidence for the opposite trend. One should bear in mind that measuring average velocities in field theory simulations is notoriously difficult (much more so than simply identifying and counting strings) so it would be premature to draw any firm conclusions. Nevertheless, this is an issue which warrants further investigation.

With these caveats in mind, we can nevertheless use the VOS model to infer, by extrapolation, the number of strings per Hubble patch and the average string velocity. As explained above, this will correspond to a choice of $N=70$ for the excess string tension. We thus find
\be
\xi_{ax}\sim4.2
\ee
\be
v_{ax}\sim0.51\,;
\ee
again we do not quote uncertainties on these for the aforementioned reasons.

\section{Discussion and conclusions}

We have used the velocity-dependent one scale (VOS) model as a framework in which to discuss several recent field theory simulations of global cosmic string networks. These networks are cosmologically relevant as a source of axions (a possible dark matter candidate), and the calculation of the amount and spectrum of axions produced by the network requires a quantitative understanding of the network's evolution. Our phenomenological model analysis confirms the presence of corrections to the canonical scale-invariant evolution, affecting both the number of strings per Hubble patch and the average string velocity and primarily depending on the excess string tension $N=\ln{(L/\delta)}$. Indications of deviations to scaling have been reported in the most recent numerical simulations \cite{Klaer,Villadoro,Kawasaki} although they remain to be fully described and quantified. 

We have attempted to calibrate the VOS model by using the simulations of \cite{Klaer} to infer the values of its free parameters: the momentum parameter, and the efficiencies of energy losses by loop production and radiation. The calibrated model satisfactorily reproduces the results of \cite{Klaer,Kawasaki}, but not those of \cite{Villadoro}. We note that apart from the fact that the various simulations have different effective $N$, their numerical algorithms and diagnostics also differ, and this may also play a role in explaining the differences.

At this point it is therefore premature to draw any firm conclusions, other than to say that exploring numerical techniques that allow the simulation of a wide range of values of $N$ is highly desirable, as are measurements of the string velocities in the simulations. We note that while the VOS model prediction for the broad dependence of the number of strings per Hubble patch seems to agree with simulations, the situation is not as clear for the velocities. It is possible that there are additional physical mechanisms, for example related to the interactions between the strings, that are dynamically important but are not included in the current model. Future higher-resolution simulations may therefore motivate improvements in the VOS model---as has happened recently in the case of domain walls \cite{Walls}.

In any case, our analysis shows how the VOS model can be used to extrapolate the results of current or future numerical simulations. These have a limited dynamic range which is compounded, for global strings, by the time-dependent tension, making it difficult to directly explore the cosmologically realistic case. While an analytic defect evolution model can only be as good as the available simulations (since it requires them for calibration), the combination of the two will enable more robust predictions of the cosmological consequences of these and other defect networks. 

\section*{Acknowledgements}

This work was inspired by discussions during the workshop {\it Cosmic Topological Defects: Dynamics and Multi-Messenger Signatures}, held at the Lorentz Center (Leiden, The Netherlands) from 22 to 26 October 2018. I warmly thank the Lorentz Center and the workshop organisers (Ana Achucarro, Leandros Perivolaropoulos and Tanmay Vachaspati) for the hospitality. Helpful comments from the referee are also acknowledged.

This work was financed by FEDER---Fundo Europeu de Desenvolvimento Regional funds through the COMPETE 2020---Operational Programme for Competitiveness and Internationalisation (POCI), and by Portuguese funds through FCT---Funda\c c\~ao para a Ci\^encia e a Tecnologia in the framework of the project POCI-01-0145-FEDER-028987.

\bibliographystyle{model1-num-names}
\bibliography{axion}

\begin{thebibliography}{16}
\expandafter\ifx\csname natexlab\endcsname\relax\def\natexlab#1{#1}\fi
\providecommand{\bibinfo}[2]{#2}
\ifx\xfnm\relax \def\xfnm[#1]{\unskip,\space#1}\fi
\bibitem[{Kibble(1976)}]{Kibble}
\bibinfo{author}{T.~W.~B. Kibble},
\newblock \bibinfo{title}{{Topology of Cosmic Domains and Strings}},
\newblock \bibinfo{journal}{J. Phys.} \bibinfo{volume}{A9}
  (\bibinfo{year}{1976}) \bibinfo{pages}{1387--1398}.
\bibitem[{Peccei and Quinn(1977)}]{Peccei}
\bibinfo{author}{R.~D. Peccei}, \bibinfo{author}{H.~R. Quinn},
\newblock \bibinfo{title}{{CP Conservation in the Presence of Instantons}},
\newblock \bibinfo{journal}{Phys. Rev. Lett.} \bibinfo{volume}{38}
  (\bibinfo{year}{1977}) \bibinfo{pages}{1440--1443}.
\bibitem[{Davis(1986)}]{Davis}
\bibinfo{author}{R.~L. Davis},
\newblock \bibinfo{title}{{Cosmic Axions from Cosmic Strings}},
\newblock \bibinfo{journal}{Phys. Lett.} \bibinfo{volume}{B180}
  (\bibinfo{year}{1986}) \bibinfo{pages}{225--230}.
\bibitem[{Klaer and Moore(2017)}]{Klaer}
\bibinfo{author}{V.~B. Klaer}, \bibinfo{author}{G.~D. Moore},
\newblock \bibinfo{title}{{How to simulate global cosmic strings with large
  string tension}},
\newblock \bibinfo{journal}{JCAP} \bibinfo{volume}{1710} (\bibinfo{year}{2017})
  \bibinfo{pages}{043}.
\bibitem[{Gorghetto et~al.(2018)Gorghetto, Hardy, and Villadoro}]{Villadoro}
\bibinfo{author}{M.~Gorghetto}, \bibinfo{author}{E.~Hardy},
  \bibinfo{author}{G.~Villadoro},
\newblock \bibinfo{title}{{Axions from Strings: the Attractive Solution}},
\newblock \bibinfo{journal}{JHEP} \bibinfo{volume}{07} (\bibinfo{year}{2018})
  \bibinfo{pages}{151}.
\bibitem[{Kawasaki et~al.(2018)Kawasaki, Sekiguchi, Yamaguchi, and
  Yokoyama}]{Kawasaki}
\bibinfo{author}{M.~Kawasaki}, \bibinfo{author}{T.~Sekiguchi},
  \bibinfo{author}{M.~Yamaguchi}, \bibinfo{author}{J.~Yokoyama},
\newblock \bibinfo{title}{{Long-term dynamics of cosmological axion strings}},
\newblock \bibinfo{journal}{PTEP} \bibinfo{volume}{2018} (\bibinfo{year}{2018})
  \bibinfo{pages}{091E01}.
\bibitem[{Martins and Shellard(1996)}]{MS1}
\bibinfo{author}{C.~J. A.~P. Martins}, \bibinfo{author}{E.~P.~S. Shellard},
\newblock \bibinfo{title}{Quantitative string evolution},
\newblock \bibinfo{journal}{Phys. Rev.} \bibinfo{volume}{D54}
  (\bibinfo{year}{1996}) \bibinfo{pages}{2535--2556}.
\bibitem[{Martins and Shellard(2002)}]{MS2}
\bibinfo{author}{C.~J. A.~P. Martins}, \bibinfo{author}{E.~P.~S. Shellard},
\newblock \bibinfo{title}{Extending the velocity-dependent one-scale string
  evolution model},
\newblock \bibinfo{journal}{Phys. Rev.} \bibinfo{volume}{D65}
  (\bibinfo{year}{2002}) \bibinfo{pages}{043514}.
\bibitem[{Fleury and Moore(2016)}]{Fleury}
\bibinfo{author}{L.~Fleury}, \bibinfo{author}{G.~D. Moore},
\newblock \bibinfo{title}{{Axion dark matter: strings and their cores}},
\newblock \bibinfo{journal}{JCAP} \bibinfo{volume}{1601} (\bibinfo{year}{2016})
  \bibinfo{pages}{004}.
\bibitem[{Martins(2016)}]{Book}
\bibinfo{author}{C.~J. A.~P. Martins}, \bibinfo{title}{Defect Evolution in
  Cosmology and Condensed Matter: Quantitative Analysis with the
  Velocity-Dependent One-Scale Model}, \bibinfo{publisher}{Springer},
  \bibinfo{year}{2016}.
\bibitem[{Martins et~al.(2004)Martins, Moore, and Shellard}]{MS3}
\bibinfo{author}{C.~J. A.~P. Martins}, \bibinfo{author}{J.~N. Moore},
  \bibinfo{author}{E.~P.~S. Shellard},
\newblock \bibinfo{title}{A unified model for vortex-string network evolution},
\newblock \bibinfo{journal}{Phys. Rev. Lett.} \bibinfo{volume}{92}
  (\bibinfo{year}{2004}) \bibinfo{pages}{251601}.
\bibitem[{Yamaguchi(2005)}]{Tension}
\bibinfo{author}{M.~Yamaguchi},
\newblock \bibinfo{title}{{Cosmological evolution of cosmic strings with time
  dependent tension}},
\newblock \bibinfo{journal}{Phys. Rev.} \bibinfo{volume}{D72}
  (\bibinfo{year}{2005}) \bibinfo{pages}{043533}.
\bibitem[{Battye and Shellard(1994)}]{Battye1}
\bibinfo{author}{R.~A. Battye}, \bibinfo{author}{E.~P.~S. Shellard},
\newblock \bibinfo{title}{{Global string radiation}},
\newblock \bibinfo{journal}{Nucl. Phys.} \bibinfo{volume}{B423}
  (\bibinfo{year}{1994}) \bibinfo{pages}{260--304}.
\bibitem[{Battye and Shellard(1996)}]{Battye2}
\bibinfo{author}{R.~A. Battye}, \bibinfo{author}{E.~P.~S. Shellard},
\newblock \bibinfo{title}{{Radiative back reaction on global strings}},
\newblock \bibinfo{journal}{Phys. Rev.} \bibinfo{volume}{D53}
  (\bibinfo{year}{1996}) \bibinfo{pages}{1811--1826}.
\bibitem[{Press et~al.(1989)Press, Ryden, and Spergel}]{PRS}
\bibinfo{author}{W.~H. Press}, \bibinfo{author}{B.~S. Ryden},
  \bibinfo{author}{D.~N. Spergel},
\newblock \bibinfo{title}{{Dynamical Evolution of Domain Walls in an Expanding
  Universe}},
\newblock \bibinfo{journal}{Astrophys. J.} \bibinfo{volume}{347}
  (\bibinfo{year}{1989}) \bibinfo{pages}{590--604}.
\bibitem[{Martins et~al.(2016)Martins, Rybak, Avgoustidis, and
  Shellard}]{Walls}
\bibinfo{author}{C.~J. A.~P. Martins}, \bibinfo{author}{I.~{\relax Yu}. Rybak},
  \bibinfo{author}{A.~Avgoustidis}, \bibinfo{author}{E.~P.~S. Shellard},
\newblock \bibinfo{title}{{Extending the velocity-dependent one-scale model for
  domain walls}},
\newblock \bibinfo{journal}{Phys. Rev.} \bibinfo{volume}{D93}
  (\bibinfo{year}{2016}) \bibinfo{pages}{043534}.

\end{thebibliography}
\end{document}